\documentclass{article}
\input{epsf}
\usepackage{amssymb}
\begin{document}

\title{The Chaplygin gas as a model for dark energy}

\author{V. Gorini,$^{1,2}$ A. Kamenshchik,$^{1,3}$
U. Moschella$^{1,2}$ \and V. Pasquier$^{4}$}

\date{}

\maketitle
\hspace{-6mm}$^{1}$Dipartimento di Scienze Matematiche, Fisiche e
Chimiche, Universit\`a dell'Insubria, \\
\hspace{2mm} Via Valleggio 11, 22100 Como, Italy\\
$^{2}$INFN sez. di Milano, Italy\\
$^{3}$L.D. Landau Institute for Theoretical Physics of Russian
Academy of Sciences, \\ \hspace{2mm}  Kosygin str. 2, 119334 Moscow, Russia\\
$^{4}$Service de Physique The\'orique,
C.E. Saclay, 91191 Gif-sur-Yvette, France\\

\abstract
{We review the essential features of the Chaplygin gas cosmological models
and provide some examples of appearance of the Chaplygin gas equation of state
 in modern physics. A possible theoretical basis for the Chaplygin gas in cosmology is discussed. The relation with scalar field and tachyon cosmological models is also considered.}

\section{Introduction}
Recent years observations of the luminosity of type Ia distant
supernovae \cite{accel1,accel2,accel3} point towards an accelerated
expansion of the universe, which implies that the pressure $p$ and
the energy density $\rho$ of the universe should violate the
strong energy condition, i.e. $\rho + 3p<0$.

The matter responsible for this condition to be satisfied at some
stage of the cosmological evolution is referred to as ``dark
energy'' (for a review see \cite{Star-Varun,Ratra-Rev,Padman-Rev}).
There are different candidates for the role of dark energy.

The most traditional candidate is a nonvanishing cosmological
constant, which can also be thought of as a perfect fluid
satisfying the equation of state $p = - \rho$. However, it remains
to understand why the observed value of the cosmological constant
is so small in comparison with the Planck mass scale.

Moreover, there also arises in this connection  the so called
"cosmic coincidence conundrum". It
amounts to the following question: why are the energy densities of
dark energy and of dust-like matter at the present epoch of the
same order of magnitude? This seems to be a problem, because it
would imply that at the time of recombination these two densities
were different by many orders of magnitude (see for
instance \cite{Varun-Rev}).

A less featureless  candidate to provide dark energy is
represented by the so called quintessence scalar field \cite{Cald}.
Scalar fields are traditionally used in inflationary models to
describe the transition from the quasi-exponential expansion of
the early universe to a power law expansion. It has been a natural
choice to try to understand the present acceleration of the
universe by also using scalar fields \cite{Star,Star1}. However, we
now deal with the opposite task, i.e. we would like to describe
the transition from a universe filled with dust-like matter to an
exponentially expanding one.

Scalar fields are not the only possibility but there are (of
course) alternatives. Among these one can point out the so called
$k$-essence models, where one deals again with a scalar field, but
with a non-standard kinetic term \cite{k-essence}. The tachyonic
models of dark energy have a similar structure, where the kinetic
term of the  tachyon field has a  form suggested  by string theory
(see the review \cite{Sen} and references therein). One can also
mention models where the role of dark energy is played by quantum
corrections to the effective action of a scalar
field \cite{Parker}.

Here we consider a recently proposed class of simple cosmological
models  based on the use of peculiar perfect fluids \cite{we}. In
the simplest case, we study the model of a universe filled with
the so called Chaplygin gas, which is a perfect fluid
characterized by the following equation of state:
\begin{equation}
p = - \frac{A}{\rho},
\label{Chapl}
\end{equation}
where  $A$ is a
positive constant.

Chaplygin introduced this equation of state \cite{origin} as a
suitable mathematical approximation for calculating the lifting
force on a wing of an airplane in aerodynamics. The same model was
rediscovered later in the same context \cite{origin1,origin2}.

The convenience of the Chaplygin gas is connected with the fact
that the corresponding Euler equations have a very large group of
symmetry, which implies their integrability. The relevant symmetry
group has been recently described in modern terms \cite{Bazeia}.

The negative pressure following from the Chaplygin equation of
state could also be used for the description of certain effects in
deformable solids \cite{solid}, of stripe states in the context of
the quantum Hall effect and of other phenomena.

It is worth mentioning a remarkable feature of the Chaplygin  gas,
namely that it has positive and bounded squared sound velocity
$$
v_s^2 = \frac{\partial p}{\partial \rho} = \frac{A}{\rho^2},
$$
which is a non-trivial fact for  fluids with negative pressure
(this follows from $\rho^2 \geq A$, see formula (\ref{solution})
below).

Beyond cosmology, the Chaplygin gas equation of state has recently
raised a growing attention \cite{Chapl} because it displays some
interesting and, in some sense, intriguingly unique features.

Indeed, Eq. (\ref{Chapl}) has a nice connection with string theory
and it can be obtained from the Nambu-Goto action for $d$-branes
moving in a $(d+2)$-dimensional spacetime in the light-cone
parametrization \cite{Chaply}. Also, the Chaplygin gas is the only
fluid which, so far, admits a supersymmetric
generalization \cite{jackiw}. We ourselves came across this
fluid \cite{we1} when studying the stabilization of branes \cite{RS}
in black hole bulks \cite{BHTZ}. An ``anti-Chaplygin'' state
equation, i.e. Eq. (\ref{Chapl}) with negative constant $A$,
arises in the description of wiggly strings \cite{Carter,Vil}.

Inspired by the fact that the Chaplygin gas possesses a negative
pressure we have undertaken the simple task of studying  a FRW
cosmology of a universe filled with this type of fluid \cite{we}.
Further theoretical developments of the model were given
in \cite{Fabris,Bilic,Bento,we2}. One of its most remarkable
properties is that it describes a transition from a decelerated
cosmological expansion to a stage of cosmic acceleration. The
inhomogeneous Chaplygin gas can do more: it is able to combine the
roles of dark energy and dark matter \cite{Bilic}.

Another model that has been discussed in some detail \cite{Bento}
is the generalized Chaplygin gas  that has two free
parameters:
\begin{equation}
p = -\frac{A}{\rho^{\alpha}},\ \ \ 0< \alpha \leq 1
\label{gen-Chap}
\end{equation}

A further possibility is to use \cite{we2} a more realistic
two-fluid cosmological model including both the Chaplygin gas and
the usual dust-like matter; this was also studied by using the
statefinder parameters \cite{statefin,statefin1}. While the model
looks less economical than pure Chaplygin,  it is more flexible
from the point of view of the comparison with observational data.
Moreover, the two-fluid model can suggest a solution of the cosmic
coincidence conundrum \cite{we2}.

The  cosmological models of the Chaplygin class have at least
three significant features: they describe a smooth transition from
a decelerated expansion of the universe to the present epoch of
cosmic acceleration; they attempt to give a unified macroscopic
phenomenological description of dark energy and dark matter; and,
finally, they represent, perhaps, the simplest deformation of
traditional $\Lambda$CDM models.

Taking into account these attractive features, it is important to
try to explain what could be the microscopic origin of the
presence of the Chaplygin gas in our universe. An interesting
attempt \cite{Bilic1} makes use of a field theory
approach to the description of a $(3+1)$-dimensional brane
immersed in a $(4+1)$-dimensional bulk \cite{Sundrum}.
Phenomenologically, the Chaplygin gas manifests itself as the
effect  of the immersion of our four-dimensional world into some
multidimensional bulk. The appearance of the Chaplygin gas in such
a context does not depend on the details of the  theory.

Another interesting feature of the Chaplygin gas is that it can be
considered as the simplest model within the family of tachyon
cosmological models. Namely, the Chaplygin gas cosmological model
can be identified with a tachyon field theoretical model with a
constant value of the field potential \cite{FKS}. Using this fact
as a starting point, we have developed a technique of construction
of tachyon models which are in general correspondence with the
present observational data, describing the contemporary epoch of
cosmic acceleration \cite{we-tac}. On the other hand, these models
have a rather rich dynamics, opening various scenarios for the
future of the universe. In particular, it is possible that the
present cosmological acceleration be followed by  a
catastrophically decelerated expansion culminating in a {\it Big
Brake} cosmological singularity. This type of cosmological
singularity is characterised by an infinite value of cosmic
deceleration which is achieved in a finite span of time.

A significant amount of work has been devoted to the comparison of
the Chaplygin cosmological predictions with observational
data$^{38-64}$. In this context, one can mention the following directions of investigation:\\
1. Supernova of type Ia observations.\\
2. Cosmic microwave background radiation.\\
3. Growth of inhomogeneities and the large scale structure of the universe.\\
4. Statistics of gravitational lensing.\\
5. $X$-ray luminosity of galaxy clusters.\\
6. Age restrictions from high-redshift objects.\\
7. New methods of diagnostics of the dark energy equation of state for future supernovae observations.\\
The restrictions coming from the studies of the large-scale
structure of the universe are crucial for the viability  of the
Chaplygin gas
models \cite{Chap-obs,Chap-obs31,Chap-obs32,Chap-obs41,Chap-obs50,Chap-obs51,Chap-obs70}.
One can safely say that a careful investigation of the non-linear
regime of the growth of inhomogeneities is necessary for the
purpose of  coming to definite conclusions concerning the
compatibility of the Chaplygin cosmologies with the observable
large-scale structure of the universe.

Finally, we remark that until recently the standard $\Lambda$CDM
model was considered as the most natural candidate for the role of
dark energy from the observational point of view. Thus, the usual
discussions of such theoretical problems as the relation between
the cosmological constant and the  Planck mass scale and the
cosmic coincidence conundrum coexisted with a tacit agreement that
there is nothing that can fit the data better than this simplest
model. One can now detect   first signs of a possible change of
the situation. An example is paper \cite{Star-change}, where some
essential arguments in favour of a non-constant ratio between the
pressure and the energy density of dark energy were put forward.
Moreover, the study of conditions under which future SNAP
(Supernova / Acceleration Probe) and
cosmic microwave background radiation observations  would be able
to rule out the simplest $\Lambda$CDM model is also attracting
attention \cite{Linde-rule}. In this context,  further study of the
Chaplygin gas cosmological model and its relatives looks
promising.

The structure of this paper is the following: in sections \ref{22}
and \ref{33}  we present some theoretical foundations of the
Chaplygin gas in modern physics;  section \ref{44} contains the
discussion of the Chaplygin cosmological model; in section
\ref{55} we consider a two-fluid cosmological model in terms of
the statefinder parameters;  section \ref{66} presents the
relationships between the Chaplygin   cosmological model and  some
corresponding scalar field and tachyon field models.

\section{Branes and the Chaplygin's equation of state} \label{22}
The rather special status of the Chaplygin fluid may be
appreciated by reviewing how its equation of state reappears in a
modern theoretical   physics context.

To this end one is led to
 consider the Nambu-Goto action for a $d$-brane moving in a
$(d+2)$-dimensional spacetime in the light-cone
parametrization \cite{Chaply}. However, to keep the discussion
elementary, we restrict our attention to the 3-dimensional
case \cite{we1} i.e. the string case. When written in the
light-cone gauge, the Hamiltonian for such a string has the
following structure:
\begin{equation}
H = \frac12 \int [\Pi^2 + (\partial_{\sigma} x)^2] d \sigma,
\label{light-cone}
\end{equation}
where $\sigma$ is a spatial world-sheet coordinate, $x$ is a
transversal spatial coordinate and $\Pi$ is its conjugate
momentum. The Hamilton equations following from (\ref{light-cone})
are very simple:
\begin{eqnarray}
&& \partial_{\tau}x = \Pi, \label{Ham1}
\\
&& \partial^2_{\tau\tau}x - \partial^2_{\sigma\sigma}x = 0. \label{Ham2}
\end{eqnarray}
At this point one wants to interpret  the  functions
\begin{eqnarray}
&& \rho(x) = (\partial_{\sigma}x)^{-1}, \label{dens-ef}
\\
&& v = \Pi, \label{vel-ef}
\end{eqnarray}
as the density and the velocity fields of a certain fluid
associated with the string.

This is substantiated by a special instance of the hodograph
transformation \cite{Chaply}, that makes one move from the
independent variables $\tau$ and $\sigma$ to the variables
$t=\tau$ and $x$. There hold the following relations:
\begin{equation}
\frac{\partial}{\partial\tau} = \frac{\partial}{\partial t} + \Pi
\frac{\partial}{\partial x} = \frac{\partial}{\partial t} + v
\frac{\partial}{\partial x}, \label{hodog}
\end{equation}
\begin{equation}
\frac{\partial}{\partial \sigma} = (\partial_{\sigma}x)\frac{\partial}{\partial x}
= \frac{1}{\rho} \frac{\partial}{\partial x}.
\label{hodog1}
\end{equation}
At this point one can easily see that the density and velocity
that we have defined always satisfy the continuity equation
\begin{equation}
\frac{\partial \rho}{\partial t} + \frac{\partial (\rho
v)}{\partial x} = 0. \label{cont}
\end{equation}

Furthermore Eqs. (\ref{hodog}) and (\ref{hodog1}) can be used to
show that  Eq. (\ref{Ham2})  for the string is equivalent to the
Euler equation
\begin{equation}
\rho\left(\frac{\partial}{\partial t} + v \frac{\partial}{\partial
x}\right)v +   \frac{\partial p}{\partial x} = 0 . \label{Euler}
\end{equation}
for the fluid, provided that the pressure field satisfies the
Chaplygin equation of state (\ref{Chapl}) with $A = 1$.

The Chaplygin  equation of state also arises in connection with
the Randall-Sundrum  model \cite{RS}. In this model one thinks of
our four dimensional Minkowski spacetime to be a brane in a higher
dimensional manifold. The difference of the Randall-Sundrum model
w.r.t. the standard Kaluza-Klein models is that the higher
dimensional manifold has now a (non-factorizable) warped structure
as given by its metric
\begin{equation}
ds^2 = e^{-2|y|/l}(dt^2 - dx_1^2 -dx_2^2 - dx_3^2) - dy^2,
\label{AdS}
\end{equation}
where $y$ is an additional fifth coordinate and
$e^{-2|y|/l}$ is the warping factor. Eq. (\ref{AdS}) is the metric
of a portion of a five-dimensional anti-de Sitter spacetime of
radius $l$.

At $y = 0$ one has the so called orbifold boundary conditions.
Here  the Christoffel symbols have finite jumps while the
components of the curvature tensor contain $\delta$-like terms. To
compensate them, one should introduce  a brane located at $y = 0$
whose tension is $\lambda = 6/l$. This tension can be  interpreted
as a nonvanishing cosmological constant on the 4-dimensional brane
spacetime, or equivalently, as a fluid living on the brane whose
state equation is $p = - \rho$. The 5-dimensional anti-de Sitter
curvature makes the graviton essentially trapped on the 4-brane.

It is possible to consider other geometries for the brane, and this in
general requires other kinds of matter on it for its stabilization.
We have considered \cite{we1} a foliation of the $(n+2)$-dimensional
anti-de Sitter spacetime by  static universes with
 topology $R \times S^n$, always imposing orbifold boundary conditions.
In this case \cite{we1} the matter on the brane is a fluid
satisfying the following state equation:
\begin{equation}
p = - \frac{(n-1)\rho}{n} - \frac{4n}{\rho l^2}.
\label{Chap-brane}
\end{equation}
In  the three-dimensional  case ($n =1$) this reduces again to the
Chaplygin gas state equation.

\section{A possible theoretical basis for the Chaplygin gas in cosmology} \label{33}

As we shall see in the next section, the Chaplygin gas cosmological
model has  interesting features. Before discussing the latter in detail,
we would like to
address the question whether there is
any fundamental mechanism to produce a Chaplygin gas source term
at the RHS of the Einstein equations.

An interesting attempt in this direction \cite{Bilic1} makes use of a $(3+1)$-brane
immersed in a $(4+1)$-bulk, following a recent stream of
ideas \cite{Sundrum}. Consider the embedding of a
$(3+1)$-dimensional brane in a $(4+1)$-dimensional bulk described
by coordinates $x^M = (x^{\mu},x^4)$, where the   index $\mu$ runs
over $0,1,2,3$. Denote the bulk metric by $g_{MN}$. Then, the
induced metric on the brane is given by
\begin{equation}
\tilde{g}_{\mu\nu} = g_{\mu\nu} - \theta_{,\mu}\theta_{,\nu},
\label{induced}
\end{equation}
where $\theta(x^{\mu})$ is a scalar field describing the embedding
of the  brane into the bulk.

The action on the brane has the following structure:
\begin{eqnarray}
&&S_{brane} =  \int d^4x \sqrt{-\tilde{g}}(-f + \cdots) = \int
d^4x \sqrt{-g}\sqrt{1 - g^{\mu\nu}\theta_{,\mu}\theta_{,\nu}}(-f +
\cdots), \label{brane-eff}
\end{eqnarray}
where the constant $f$ gives a brane tension and  the dots
$\cdots$ stay for other possible contributions. Equation
(\ref{brane-eff}) follows from the identity
\[\det(a_{ij} - b_ib_j) = \det(a_{ij}) \,(1 - b_m (a^{-1})_{mn} b_n),\]
whose proof is straightforward.

The energy-momentum tensor following from the tension-like
 action (\ref{brane-eff}) is then
\begin{equation}
T_{\mu\nu} = f\left(\frac{\theta_{,\mu}\theta_{,\nu}}{\sqrt{1 -
g^{\mu\nu}\theta_{,\mu}\theta_{,\nu}}} + g_{\mu\nu} \sqrt{1 -
g^{\mu\nu}\theta_{,\mu}\theta_{,\nu}}\right). \label{tens-ef}
\end{equation}
This expression corresponds to a perfect fluid energy-momentum
tensor
\[T_{\mu\nu} = (\rho + p)u_{\mu}u_{\nu} - p g_{\mu\nu},\]
provided one makes the following identifications: the
four-velocity $$u_{\mu} =
\frac{\theta_{,\mu}}{\sqrt{g^{\mu\nu}\theta_{,\mu}\theta_{,\nu}}};$$
the pressure and the energy density:
\begin{equation}
p = -f \sqrt{1 - g^{\mu\nu}\theta_{,\mu}\theta_{,\nu}}
\label{pressure-ef}
\end{equation}
\begin{equation}
\rho = f \frac{1}{\sqrt{1 -
g^{\mu\nu}\theta_{,\mu}\theta_{,\nu}}}. \label{energy-ef}
\end{equation}
It follows that the pressure and energy density exactly satisfy
the Chaplygin's equation of state (\ref{Chapl}) with $A = f^2$.

\section{FRW cosmology with the Chaplygin gas} \label{44}
We consider a homogeneous and isotropic universe with the metric
\begin{equation}
ds^2 = dt^2 - a^2(t)dl^2, \label{Fried}
\end{equation}
where $dl^2$ is the metric of a 3-manifold of constant curvature
($K= 0,\pm 1$), and the expansion factor $a(t)$ evolves according
to the Friedmann equation
\begin{equation}
\frac{\dot{a}^2}{a^2} = \rho - \frac{K}{a^2}. \label{Fried1}
\end{equation}
Energy conservation
\begin{equation}
d(\rho \, a^3) = - p\ d (a^3) \label{Fried2}
\end{equation}
 together with the equation of state (\ref{Chapl}) give the following relation:
\begin{equation}
\rho = \sqrt{A + \frac{B}{a^6}},
\label{solution}
\end{equation}
where $B$ is an integration constant. By choosing a positive value
for $B$ we see that for small $a$ (i.e. $a^6 \ll B/A$)  the
expression (\ref{solution}) is approximated by
\begin{equation}
\rho \sim \frac{\sqrt{B}}{a^3} \label{dust}
\end{equation}
which corresponds to  a universe dominated by dust-like matter. For
large values of the cosmological radius $a$ it follows that
\begin{equation}
\rho \sim \sqrt{A}, \ p \sim - \sqrt{A},
\label{cosm}
\end{equation}
which, in turn, corresponds to an empty universe with a
cosmological constant $\sqrt{A}$ (i.e a de Sitter universe). In
the flat case it is also possible  to find the exact solution as
follows:
\begin{eqnarray}
&& t = \frac{1}{6^4\!\!\sqrt{A}}\left(\ln
\frac{^4\!\!\sqrt{A+\frac{B}{a^6}}+^4\!\!\sqrt{A}}
{^4\!\!\sqrt{A+\frac{B}{a^6}}-^4\!\!\sqrt{A}} - 2\arctan
^4\!\!\sqrt{1+\frac{B}{Aa^6}} + \pi\right).
\label{exact}
\end{eqnarray}

Note that $\rho = \sqrt A$ solves the equation
\begin{equation}
\rho +
p=\rho -\frac{A}\rho = 0.
\end{equation}
The circumstance that
this equation has a nonzero solution lies at the heart of the
possibility of interpreting the model as a ``quintessential''
one. Let us estimate the
constant $A$ by comparing our expressions for pressure and energy
with observational data. An indirect and naive way to do it is to
consider the nowadays accepted values for the contributions of
matter and cosmological constant to the energy density of the
universe. To use these data we decompose pressure and energy
density as follows:
\begin{eqnarray}
&& p = p_{\Lambda} + p_{M} = -\Lambda, \\
&& \rho = \rho_{\Lambda} + \rho_{M} = \Lambda + \rho_{M}.
\end{eqnarray}
An application of Eq. (\ref{Chapl}) gives
\begin{equation}
A = \Lambda (\Lambda + {\rho_{M}}). \label{param}
\end{equation}
If the cosmological constant contributes seventy percent to the
total energy we get $\sqrt{A} \approx 1.2 \,\Lambda$. We now observe
that, in the context   of a Chaplygin cosmology, once an expanding
universe starts accelerating it cannot decelerate any more. Indeed
Eqs. (\ref{Fried1}) and (\ref{Fried2}) imply that
\begin{equation}
\frac{\ddot{a}}{a} = -\frac12 (\rho + 3p). \label{Fried3}
\end{equation}
Condition $\ddot a>0$ is equivalent to
\begin{equation}
a^6 > \frac{B}{2A}, \label{cond}
\end{equation}
which is obviously preserved by time evolution in an expanding
universe. It thus follows that the observed value of the
(effective) cosmological constant  will increase up to $1.2 \,
\Lambda$.

There is a relation of our Chaplygin cosmology with cosmologies
based on fluids admitting a bulk viscosity proportional to a power
of the density \cite{barrow1,barrow2,barrow3,barrow4}. In the flat
$K=0$ case the FRW equations for Chaplygin fit in this scheme as a
special case \cite{barrow2} and indeed a transition from power law
to exponential expansion was already noticed \cite{barrow1,Ellis}.
However, since the state equation for the corresponding fluid is
different from our Eq. (\ref{Chapl}) this coincidence of the
solutions is destroyed by any small perturbation, for instance by
a small spatial curvature or by adding another matter source.

Considering now the subleading terms in  Eq. (\ref{solution}) at
large values of $a$ (i.e. $a^6 \gg B/A$), one obtains the
following expressions for the energy and pressure:
\begin{eqnarray}
&& \rho \approx \sqrt{A} + \sqrt{\frac{B} {4 A}}\ a^{-6},
\label{stiff}\\
&&p \approx -\sqrt{A} + \sqrt{\frac{B} {4 A}}\ a^{-6}.
\label{stiffpr}
\end{eqnarray}
Eqs. (\ref{stiff}) and (\ref{stiffpr}) describe the mixture of a
cosmological constant $\sqrt{A}$ with a type of matter known as
``stiff'' matter,  described by the following equation of
state:
\begin{equation}
p = \rho.
\label{stiffeq}
\end{equation}
Note that a massless scalar field is a particular instance  of
stiff matter. Therefore, in a generic situation, a Chaplygin
cosmology can be looked upon as interpolating between different
phases of the universe: from a dust dominated universe to a de
Sitter one passing through an intermediate phase which is
the  mixture just mentioned above. The interesting point, however,
is that such an evolution is accounted for by using  one fluid only.

For open or flat Chaplygin cosmologies ($K=-1,0$), the universe
always evolves from a decelerating to an accelerating epoch.  For
the closed Chaplygin cosmological models ($K=1$), the Friedmann
equations (\ref{Fried1}) and (\ref{Fried3}) tell us that it is
possible to have a static Einstein universe solution $ a_0 =
{(3A)}^{-\frac{1}{4}}$ provided the following condition holds:
\begin{equation}
B = \frac{2}{3\sqrt{3A}} \label{Einstein}.
\end{equation}
When $ B > \frac{2}{3\sqrt{3A}} $ the cosmological radius $a(t)$
can take any value,  while if $$ B < \frac{2}{3\sqrt{3A}} $$ there
are two possibilities: either
\begin{equation}
a< a_1 = \frac{1}{\sqrt{3A}} \left(\sqrt{3}\sin\frac{\varphi}{3} -
\cos\frac{\varphi}{3}\right) \label{amax}
\end{equation}
or
\begin{equation}
a> a_2 = \frac{2}{\sqrt{3A}}\cos\frac{\varphi}{3}, \label{amin}
\end{equation}
where $\varphi = \pi - \arccos  {3\sqrt{3A}B}/2 $. The region
 $a_1 < a < a_2$ is not accessible.
Further information on the dynamics of the Chaplygin gas
cosmological model can be found in the literature \cite{Khal,Mar}.

For the generalized Chaplygin gas \cite{we,Bento} the dependence of
the energy density on the cosmological radius is
\begin{equation}
\rho = \left(A + \frac{B}{a^{3(1+\alpha)}}\right)^{\frac{1}{1+\alpha}}.
\label{gen-Chap1}
\end{equation}
This type of matter at the beginning of the cosmological evolution
behaves like dust and at the end of the evolution like a
cosmological constant, while during the intermediate stage it
could be treated as a mixture of two-fluids: the cosmological
constant and a perfect fluid with equation of state $p = \alpha
\rho$. The generalized Chaplygin gas cosmological models have an
additional free parameter $\alpha$ to play with and are convenient
for the comparison with observational data. However, the prospects
of the construction of a physical theory explaining the origin of
these models seem even less evident than those for the true
Chaplygin gas with $\alpha = 1$.

\section{The statefinder parameters and a two-fluid cosmological
model}\label{55} Since  models trying to provide a description (if
not an explanation) of the cosmic acceleration are proliferating,
there exists the problem of discriminating between the various
contenders. To this aim a new proposal introduced
in \cite{statefin} may turn out useful, which exploits   a pair of
parameters $\{r,s\}$, called ``statefinder''. The relevant
definition is as follows:
\begin{equation}
r \equiv \frac{\stackrel{\cdots}{a}}{a H^3},\ \ s \equiv \frac{r-1}{3(q-1/2)},
\label{statefinder}
\end{equation}
where $ H \equiv \frac{\dot{a}}{a}$ is the Hubble constant  and $
q \equiv -\frac{\ddot{a}}{a H^2} $ is the deceleration parameter.
The new feature of the statefinder is that it involves the third
derivative of the cosmological radius.

Trajectories in the $\{s,r\}$-plane corresponding to different
cosmological models exhibit qualitatively different behaviours.
$\Lambda$CDM model diagrams correspond to the fixed point $s=0$,
$r=1$. The so-called "quiessence" models \cite{statefin} are
described by vertical segments with $r$ decreasing from $r=1$
down to some definite value. Tracker models \cite{tracker,tracker1} have
typical trajectories similar to arcs of parabola lying in the
positive quadrant with positive second derivative.

The current location of the parameters $s$ and $r$ in these
diagrams can be calculated in models (given the deceleration
parameter); it may also be extracted from data coming from future SNAP
(SuperNovae Acceleration Probe)-type experiments
 \cite{statefin}. Therefore, the statefinder diagnostic combined
with  forthcoming SNAP observations may possibly be used to discriminate
among different dark energy models.

Here, we consider
the one-fluid pure Chaplygin gas model and a two-fluid model
where dust is also present \cite{we2}.
We show that these models are
different from those considered in \cite{statefin}.

To begin with, let us rewrite the formulae for the statefinder
parameters in a form convenient for our
purposes.
Since
\begin{equation}
\dot{p} = \frac{\partial p}{\partial \rho}\, \dot{\rho} =
-3\sqrt{\rho}\, (\rho + p) \frac{\partial p}{\partial \rho}\, ,
\label{pressure1}
\end{equation}
we easily get:
\begin{equation}
r = 1 + \frac{9}{2}\left(1 + \frac{p}{\rho}\right) \frac{\partial
p}{\partial \rho} \label{s-find1}\, , \;\;\;\; s = \left(1 +
\frac{\rho}{p}\right) \frac{\partial p}{\partial \rho}\, .
\end{equation}
For the Chaplygin gas one has simply that
\begin{equation}
v^2_s = \frac{\partial p}{\partial \rho}= \frac{A}{\rho^2} =
-\frac{p}{\rho} = 1+s \label{x-define}
\end{equation}
and therefore
\begin{equation}
r = 1 - \frac{9}{2}\, s\, (1+s). \label{r-find2}
\end{equation}
Thus, the curve $r(s)$ is an arc of parabola.
It is easy to see that
\begin{equation}
v^2_s = \frac{A}{A + \frac{B}{a^6}}.
\label{x-define0}
\end{equation}
When the cosmological scale factor $a$ varies from $0$ to
$\infty$ the velocity of sound varies from $0$ to $1$ and $s$
varies from $-1$ to $0$. Thus in our model the statefinder $s$
takes negative values; this feature is not shared by quiessence
and tracker models \cite{statefin}.

\begin{figure}[h]
  \centering
\epsfxsize 9cm \epsfbox{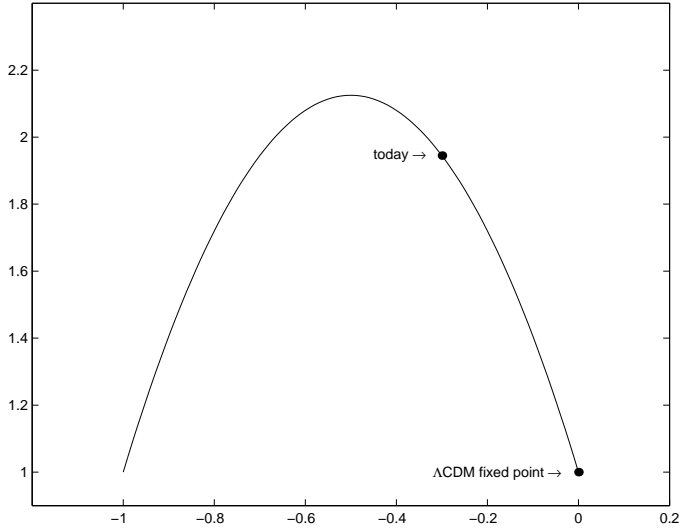}
  \caption{s-r evolution diagram for the pure Chaplygin gas}\label{Fig.1}
\end{figure}

As $s$ varies in the interval $[-1,0]$, $r$ first increases from
$r=1$ to its maximum value and then decreases  to the
$\Lambda$CDM fixed point $s=0$, $r=1$ (see Fig. \ref{Fig.1}).

\noindent If $q \approx - 0.5$ the current values of the
statefinder (within our model) are $ s \approx -0.3,\ \ r \approx
1.9 $. In \cite{statefin}  an interesting
numerical experiment based on 1000 realizations of a SNAP-type
experiment, probing a fiducial $\Lambda$CDM model is reported. Our values of
the statefinder lie outside the three-sigma confidence region
displayed in \cite{statefin}. Based on this fact it can be
expected that future SNAP experiments should be able to
discriminate between the pure Chaplygin gas model and the standard
$\Lambda$CDM model.

\begin{figure}[h]
  \centering
\epsfxsize 8cm
\epsfbox{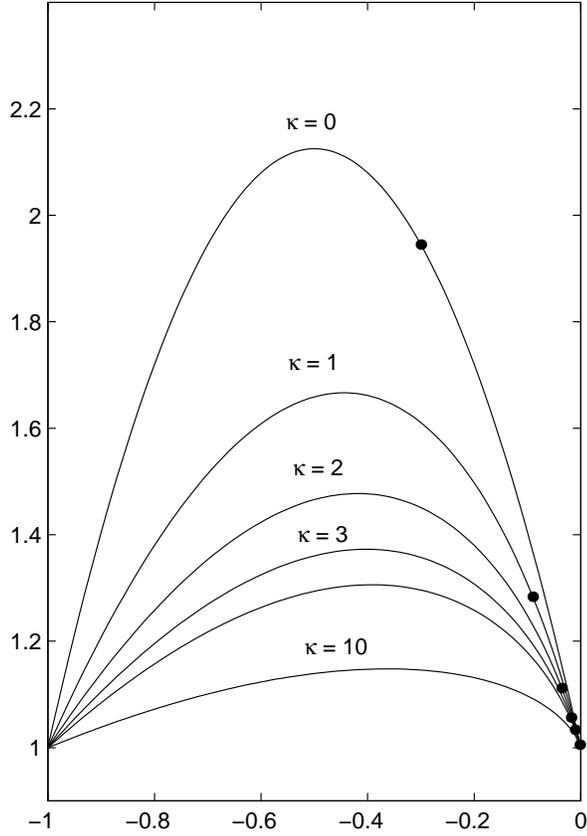}
  \caption{s-r evolution diagram for the Chaplygin gas mixed with dust.
  Dots locate the current value of  the statefinder}\label{Fig.2}
\end{figure}

Now consider a more "realistic" cosmological model which,
besides a Chaplygin's component, contains also a dust component.
For a two-component fluid Eqs. (\ref{s-find1}) take the following
form:
\begin{equation}
r = 1 + \frac{9}{2({\rho+ \rho_1})}\left[{ \frac{\partial
p}{\partial \rho}(\rho + p)+\frac{\partial p_1}{\partial
\rho_1}(\rho_1 + p_1) }\right], \label{r-find3}
\end{equation}
\begin{equation}
s = \frac{1}{p + p_1} \left[\frac{\partial p}{\partial \rho}(\rho
+ p) + \frac{\partial p_1}{\partial \rho_1}(\rho_1 + p_1)\right].
\label{s-find3}
\end{equation}
If one of the fluids is dust, i.e. $p_1 = p_d = 0$, the above
formulae become
\begin{equation}
r = 1 + \frac{9(\rho + p)}{2(\rho + \rho_d)} \frac{\partial
p}{\partial \rho}\, ,\;\;\;\; s = \frac{\rho + p}{p}
\,\frac{\partial p}{\partial \rho}. \label{s-find4}
\end{equation}
If the second fluid is the Chaplygin gas, proceeding exactly as
before we obtain the following relation:
\begin{equation} r = 1 -
\frac{9}{2}\frac{s(s+1)}{1 + \frac{\rho_d}{\rho}}. \label{r-find5}
\end{equation}
To find the term $\rho_d/\rho$ we write down the dependence of the
dust density on the cosmological scale factor:
\begin{equation}
\rho_d = \frac{C}{a^3}, \label{dust1}
\end{equation}
where $C$ is a positive constant. Eq. (\ref{x-define0}) gives
$A a^6 + B = -\frac{B}{s}$ and therefore
\begin{equation}
\frac{\rho_d}{\rho} = \frac{C}{\sqrt{A a^6 + B}} =  \kappa
\sqrt{-s} , \label{dens-rel2}
\end{equation}
where the constant $ \kappa = {C}/{\sqrt{B}} $ is the ratio
between the energy densities of dust and of the Chaplygin gas at
the beginning of the cosmological evolution. Thus
\begin{equation}
r = 1 - \frac{9}{2}\frac{s(s+1)}{1 + \kappa\sqrt{-s}}.
\label{r-find6}
\end{equation}
Graphs of the function (\ref{r-find6}) for different choices of
$\kappa$ are plotted in Fig. \ref{Fig.2}.

In this case there are choices of the parameters so that the
current values of the statefinder are close to the $\Lambda$CDM
fixed point. For $\kappa = 1$ we have $s=-0.09$ and $r=1.2835$; by
increasing $\kappa$ we get closer and closer to the point
$(0,1)$. Already for $\kappa = 2$ we get $s=0.035$, $r=1.11$
while for $\kappa \gtrsim 5$ the statefinder essentially coincides
with the $\Lambda$CDM fixed point (see Fig. \ref{Fig.2}).

Thus our  two-fluid cosmological models (with  $\kappa$ say
bigger than 5) cannot be discriminated from the $\Lambda$CDM model
on the basis of the statefinder analysis.

However, even if the Chaplygin component closely mimics today the
cosmological constant, this neither spoils the interest of the
two-fluid model nor makes it equivalent to $\Lambda$CDM; for
instance, one advantage of the model is that it may suggest a
solution to the cosmic coincidence conundrum: here the initial
values of the energies of dust and of the Chaplygin gas can be of
the same order of magnitude. In particular the value $\kappa = 1$
is not excluded by current observations. This may be seen  by
using the results of  \cite{Chap-obs} and taking into account the
relation
\begin{equation}
\kappa = \frac{\Omega_m}{(1-\Omega_m)\sqrt{1-v^2_s}},
\label{kappa}
\end{equation}
where $\Omega_m = \frac{\rho_d}{\rho_d+\rho}$ and where $\rho,
\rho_d$ and $v_s$ are evaluated at the present epoch.

Further details concerning application of the statefinder diagnostic to
the study of Chaplygin gas models can be found in \cite{we2,statefin1}.

\section{The Chaplygin gas, scalar fields, tachyons and the future of the
universe}\label{66} It is well-known that for isotropic
cosmological models, given the dependence of the cosmological
radius on time it is always possible to construct a potential for
a minimally coupled scalar field model, which would reproduce this
cosmological evolution (see e.g. \cite{Star}), provided rather
general reasonable conditions are satisfied. Sometimes, it is
possible to construct an explicit  scalar field potential, which,
provided some special initial conditions are chosen, can reproduce
the  evolution arising in some perfect fluid cosmological
model \cite{barrow3,we-tac}.

Consider the Lagrangian
\begin{equation}
L(\phi)= \frac{1}{2}\dot{\phi}^2 - V(\phi) \label{scalar}
\end{equation}
and set the energy density  of the field equal to that of the
Chaplygin gas:
\begin{equation}
\rho_\phi = \frac{1}{2}\dot{\phi}^2 + V(\phi) = \sqrt{A +
\frac{B}{a^6}}. \label{scalar1}
\end{equation}
The corresponding "pressure" coincides with the Lagrangian
density:
\begin{equation}
p_\phi = \frac{1}{2}\dot{\phi}^2 - V(\phi) = -\frac{A}{\sqrt{A +
\frac{B}{a^6}}}. \label{scalar2}
\end{equation}
It immediately follows that
\begin{equation}
\dot{\phi}^2 = \frac{B}{a^6 \sqrt{A + \frac{B}{a^6}}}
\label{scalar3}
\end{equation}
and
\begin{equation}
V(\phi) = \frac{2 a^6 \left(A + \frac{B}{a^6}\right) - B}{2
a^6\sqrt{A + \frac{B}{a^6}}}. \label{scalar4}
\end{equation}
We restrict ourselves to the flat case $K=0$. Then
 Eq. (\ref{scalar3}) also implies that
\begin{equation}
\phi' = \frac{\sqrt{B}}{a(A a^6 + B)^{1/2}}, \label{scalar5}
\end{equation}
where prime means differentiation w.r.t. $a$. This  equation can
be integrated and it follows that
\begin{equation}
a^6 = \frac{4B\exp(6\phi)}{A (1 - \exp(6\phi))^2}. \label{scalar6}
\end{equation}
Finally, by substituting the latter expression for the
cosmological radius in Eq. (\ref{scalar4}) one obtains the
following potential, which has a surprisingly simple form:
\begin{equation}
V(\phi) = \frac{1}{2}\sqrt{A}\left( \cosh 3\phi + \frac{1}{\cosh
3\phi}\right). \label{potential}
\end{equation}
Note that the potential does not depend on the integration
constant $B$ and therefore it reflects only the state equation
(\ref{Chapl}) as it should.

The cosmological evolution of the model with a scalar field with  potential (\ref{potential}) coincides with that of the Chaplygin gas model provided the initial values $\phi(t_0)$ and $\dot{\phi}(t_0)$ satisfy the relation
$\dot{\phi}^4(t_0) = 4(V^2(\phi(t_0)) - A)$.

We have mentioned in the Introduction, that one of the most popular candidates for the role of dark energy is a tachyon field described by the effective action \cite{Sen1,Garousi}
\begin{equation}
S = -\int d^4x \sqrt{-g}V(T)\sqrt{1 - g^{\mu\nu}T_{,\mu}T_{,\nu}},
\label{tac-ac}
\end{equation}
where $V(T)$ is a tachyon potential.
For a spatially homogeneous model
 \[L = -V(T)\sqrt{1-\dot{T}^2}.\]
The energy density of the tachyon field is
\begin{equation}
\rho = \frac{V(T)}{\sqrt{1-\dot{T}^2}}
\label{en-tac}
\end{equation}
while the pressure is
\begin{equation}
p = - V(T)\sqrt{1-\dot{T}^2}.
\label{pres-tac}
\end{equation}
It is easy to see \cite{FKS} that for the constant tachyon potential $V(T) = V_0$ the pressure (\ref{pres-tac}) and the energy density (\ref{en-tac}) are connected by the Chaplygin state equation (\ref{Chapl}) with $A = V_0^2$.

Thus, we see that the Chaplygin gas model coincides with the simplest tachyon cosmological model. It is interesting to construct also other tachyon potentials
reproducing the dynamics of some perfect fluid models.
For example \cite{Feinstein,Padman} a tachyon model with a potential
\begin{equation}
V(T) = \frac{4\sqrt{-k}}{9(1+k)T^2},\ \ -1 < k < 0
\label{FP}
\end{equation}
has a solution
\begin{equation}
T = \sqrt{1+k} t,
\label{solution1}
\end{equation}
corresponding to the cosmological evolution
\begin{equation}
a = a_0 t^{\frac{2}{3(1+k)}},
\label{power-law}
\end{equation}
which, in turn, could be obtained in the model with a perfect fluid obeying the state equation
\begin{equation}
p = k \rho.
\label{FP1}
\end{equation}
It is not difficult to show \cite{we-tac}, that for $k > 0$ it is impossible to reproduce the power-law cosmological evolution (\ref{power-law}) using a tachyon action. Instead, one can use a ``pseudo'' - tachyon action with the Lagrangian
\begin{equation}
L = V(T) \sqrt{\dot{T}^2 - 1}
\label{pseudo}
\end{equation}
with the potential
\[V(T) = \frac{4\sqrt{k}}{9(1+k)T^2}.\]
In this case the corresponding solution (\ref{solution1}) preserves its form.

We have considered \cite{we-tac} a more complicated toy tachyon model. Studying  a two-fluid cosmological model, where one of the fluids is the cosmological constant and the other fluid obeys the state equation $p = k \rho, -1 < k < 1$, one gets the following expression for the cosmological evolution
\begin{equation}
a(t) = a_0 \left(\sinh \frac{3\sqrt{\Lambda}(1+k)t}{2}\right)^{\frac{2}{3(1+k)}}.
\label{toy-evol}
\end{equation}
The same evolution can be reproduced in the tachyon model with a potential
\begin{equation}
V(T) = \frac{\Lambda}{\sin^2\left(\frac{3\sqrt{\Lambda(1+k)}T}{2}\right)}
\sqrt{1 - (1+k)\cos^2\left(\frac{3\sqrt{\Lambda(1+k)}T}{2}\right)}.
\label{pot-toy}
\end{equation}
The solution of the tachyon equation of motion corresponding to the evolution
(\ref{toy-evol}) has the form
\begin{equation}
T(t) = \frac{2}{3\sqrt{\lambda(1+k)}}\arctan
\sinh \frac{3\sqrt{\Lambda}(1+k)t}{2}
\label{sol-toy}
\end{equation}
and could be obtained provided some special initial conditions are chosen.

Considering all possible initial conditions we get a  rich family of cosmological evolutions, which are rather different from (\ref{toy-evol}), representing a simple two-fluid model \cite{we-tac}.
Here we encounter two ``surprises''. First, when $k > 0$, for the description of the dynamics of the model it is necessary to consider regions of the phase plane $(T,\dot{T})$, where the tachyon action (\ref{tac-ac}) with the potential (\ref{pot-toy}) is not well-defined and should be substituted by a pseudo-tachyon action. Second, (again for the case $k > 0$) there are two types of trajectories : \\a) infinitely expanding universes; \\b) universes, hitting
a cosmological singularity of a special type which we call {\it Big Brake}, and which is characterised by the following behaviour of the cosmological radius
\begin{equation}
\ddot{a}(t_B) = -\infty,\ \dot{a}(t_B) = 0,\ 0 < a(t_B) < \infty.
\label{brake}
\end{equation}
Here $t_B$ means the final moment of time when the Big Brake is achieved.

In the current literature devoted to the future of the universe
the following scenarios are usually condsidered:\\
a) an infinite asymptotically de Sitter expansion,\\
b) present accelerated expansion followed by contraction and the achieving of a Big Crunch cosmological singularity (see, e.g. \cite{Kal-Lin1,ASS})\\
c) an accelerated expansion culminating in a Big Rip cosmological singularity arising in the phantom dark energy cosmological models (see, e.g. \cite{phant,phant1,phant2}). At the  Big Rip singularity the cosmological radius
and the Hubble parameter achieve an infinite value in a finite interval of time.

On the basis of the above considerations,
it seems reasonable to envisage an additional scenario,
in which the present  cosmic acceleration is followed by a decelerated expansion culminating in the hitting of a Big Brake cosmological singularity. The latter, as we have seen, can be described in terms of a rather simple tachyon model \cite{we-tac}.

\section*{Acknowledgments}
A.K. is grateful to CARIPLO Science Foundation and to University
of Insubria for financial support. His work was also partially
supported by the Russian Foundation for Basic Research under the
grant No 02-02-16817 and by the scientific school grant No. 2338.2003.2 of the Russian Ministry of Science and Technology.

\end{document}